%% file: root.tex
\documentclass[conference]{IEEEtran}
\IEEEoverridecommandlockouts
\usepackage{caption}

\usepackage{cite}
\usepackage{amsmath,amssymb,amsfonts,amsthm}
\usepackage{algorithmic}
\usepackage{graphicx}
\usepackage{svg}
\usepackage{textcomp}
\usepackage{xcolor}
\usepackage{xfrac}
\usepackage{comment}
\usepackage{booktabs} 
\usepackage{multirow} 
\usepackage{xurl}
\usepackage{bm}
\usepackage{acro}
\usepackage{tikz}
\usetikzlibrary{shapes.geometric, arrows, positioning, fit}
\usepackage[dvipsnames]{xcolor}
\usepackage{tabularx}
\usepackage{subcaption} 
\usepackage{siunitx}
\usepackage{algorithm}
\usepackage{algorithmic}
\usepackage{nicefrac}
\usepackage{bm}
\usepackage{svg}

\input{src/acro}

\usepackage[hidelinks]{hyperref}
\def\BibTeX{{\rm B\kern-.05em{\sc i\kern-.025em b}\kern-.08em
    T\kern-.1667em\lower.7ex\hbox{E}\kern-.125emX}}

\PassOptionsToPackage{svgnames}{xcolor}
\usepackage{tcolorbox}
\tcbuselibrary{skins,breakable}
\usetikzlibrary{shadings,shadows}
    {\endtcolorbox}

\begin{document}
\bstctlcite{BSTcontrol} 

\UseRawInputEncoding
\title{From Big Data to Fast Data: Towards High-Quality Datasets for Machine Learning Applications from Closed-Loop Data Collection}
\author{\IEEEauthorblockN{Philipp Reis, Jacqueline Henle, Stefan Otten  and Eric Sax}
\IEEEauthorblockA{
FZI Research Center for Information Technology, Karlsruhe, Germany\\
Email: \{reis, henle, otten, sax\}@fzi.de}
}
\theoremstyle{definition}
\newtheorem{definition}{Definition}[section]

\maketitle
\bibliographystyle{IEEEtran}
\input{src/00_abstract}

\input{src/01_Introduction}

\input{src/02_Problem_Statement}
\input{src/03_datasets}

\input{src/04_strategies}
\input{src/05_Fast_Data}
\input{src/06_Outlook}

\bibliography{references.bib}
\end{document}

%% file: src/acro.tex
\DeclareAcronym{ml}{
    short = ML, 
    long = Machine Learning
}

\DeclareAcronym{ai}{
    short = AI, 
    long = Artificial Intelligence
}

\DeclareAcronym{odd}{
    short = ODD, 
    long = Operational Design Domain
}

%% file: src/00_abstract.tex
\begin{abstract}

The increasing capabilities of machine learning models, such as vision-language and multimodal language models, are placing growing demands on data in automotive systems engineering, making the quality and relevance of collected data enablers for the development and validation of such systems. Traditional Big Data approaches focus on large-scale data collection and offline processing, while Smart Data approaches improve data selection strategies but still rely on centralized and offline post-processing.

This paper introduces the concept of Fast Data for automotive systems engineering. The approach shifts data selection and recording onto the vehicle as the data source. By enabling real-time, context-aware decisions on whether and which data should be recorded, data collection can be directly aligned with data quality objectives and collection strategies within a closed-loop. This results in datasets with higher relevance, improved coverage of critical scenarios, and increased information density, while at the same time reducing irrelevant data and associated costs. The proposed approach provides a structured foundation for designing data collection strategies that are aligned with the needs of modern machine learning algorithms. It supports efficient data acquisition and contributes to scalable and cost-effective ML development processes in automotive systems engineering.

\end{abstract}

%% file: src/01_Introduction.tex
\section{Introduction}

Real-world vehicle data is a key enabler for the development and evaluation of automotive functions. This is particularly important for machine learning-based systems, whose performance and safety depend strongly on the quality, representativeness, and relevance of the underlying datasets. Consequently, development processes in automotive systems engineering are increasingly shaped by data-driven approaches~\cite{bach_data-driven_2017}, in which data collection, curation, and evaluation become integral parts of the overall systems engineering process~\cite{petersen_towards_22}.

Historically, automotive data collection has largely followed a \textit{Big Data} paradigm, in which large volumes of data are recorded first and evaluated later in centralized backend infrastructures. In this setting, relevant information extraction, curation, and analytics are predominantly performed offline after data transmission and storage. However, driving data is highly uneven in its informational value: common scenarios are massively overrepresented, while novel, rare, or safety-critical situations occur only sparsely. As a result, exhaustive recording leads to substantial redundancy, delayed value extraction, and high storage, transmission, and data handling costs.

\begin{figure}
    \centering
    \includegraphics[width=0.7\linewidth]{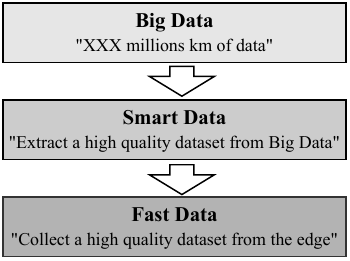}
    \caption{Shift from Big Data to Fast Data.}
    \label{fig:FastData}
\end{figure} 

\textit{Smart Data} approaches address this limitation by selecting and filtering relevant subsets of data based on their estimated utility. Nevertheless, these approaches still predominantly rely on centralized, post-hoc evaluation and therefore require large amounts of data to be recorded and transmitted before their relevance can be assessed. This maintains long feedback cycles between data generation, dataset curation, and model improvement.

To address these limitations, we enhance the concept of \textit{Fast Data}~\cite{miloslavskaya_big_2016} for AI systems engineering, shifting data selection and recording control to the edge, i.e., to the vehicle itself (cf.~Figure~\ref{fig:FastData}). In this paradigm, decisions about whether and how data should be recorded are made at the time of generation, based on the current system context as well as predefined collection objectives and strategies~\cite{reis2025datadrivennoveltyscorediverse}. Thus, data collection becomes part of a closed feedback loop that continuously aligns observed system behavior with dataset needs. This reduces the latency between data generation and its use in model development, while improving resource efficiency by avoiding the storage, transmission, and processing of irrelevant data.

\input{fig/comp_big_smart_fast}

The dimensions in Table \ref{tab:data_paradigms} characterize how data collection strategies differ with respect to where and when selection decisions are taken, how strongly these decisions depend on the dataset, and how adaptively the collection process is controlled. In this progression, data collection evolves from dataset-agnostic and retrospective logging toward dataset-specific and feedback-driven collection. The control strategy determines whether decisions are taken by static rules, delayed feedback, or a continuously updated closed loop.

Together, these dimensions illustrate the transition from volume-driven to purpose-driven data collection in the Fast Data paradigm, where data collection is no longer a passive logging task but an actively controlled procedure.


\subsection{Contribution}

Motivated by the shift toward Fast Data, this work investigates how real-world vehicle data collection can be aligned with the requirements of high-quality machine learning datasets. In particular, the following research questions are addressed:

\begin{itemize}
    \item \textbf{RQ1}: What characterizes a high-quality dataset in the context of data collection within the Fast Data paradigm?
    \item \textbf{RQ2}: Which data collection strategies and architectural principles enable the acquisition of high-quality data under Fast Data constraints?
\end{itemize}

To address these questions, this paper makes four contributions:

\begin{itemize}
    \item Derivation of a {dataset quality model} for Fast Data-driven data collection, identifying the key quality characteristics that determine dataset utility in real-world vehicle applications.

    \item Analysis of alternative pathways of dataset generation, showing why selective real-world data collection remains indispensable despite complementary advances in synthetic and pre-existing data sources.

    \item Development of a taxonomy of current data collection strategies by distinguishing data-intrinsic from dataset-specific triggers and by analyzing their limitations with respect to dataset quality objectives.

    \item Formulation of Fast Data collection as a closed-loop, dataset-state-aware, and structure-aware paradigm, together with the derivation of the core requirements of a corresponding collection architecture for high-quality dataset generation under source-side resource constraints.
\end{itemize}

%% file: fig/comp_big_smart_fast.tex
 \begin{table}[h]
\caption{Comparison of data collection paradigms in automotive systems engineering}
\label{tab:data_paradigms}
\centering
\resizebox{\linewidth}{!}{
\begin{tabular}{lccc}
\hline
\textbf{Dimension} & \textbf{Big Data} & \textbf{Smart Data} & \textbf{Fast Data} \\[1pt]
\hline
Decision location   & Backend        & Backend         & Vehicle \\
Selection timing    & Post-hoc     & Post-hoc      & Real-time \\
Dataset relation  & Dataset-agnostic & Dataset-conditioned & Dataset-specific \\
Adaptivity          & Static       & Partial     & Adaptive \\
Control principle   & Open-loop    & Delayed Closed-loop      & Closed-loop \\
\hline
\end{tabular}
}
\end{table}

%% file: src/02_Problem_Statement.tex
\section{Problem Formulation}
\label{sec:problem_formulation}

In classical knowledge discovery in databases (KDD), data analysis assumes the existence of a data warehouse from which a target dataset can be selected and curated offline. In the Fast Data paradigm for vehicular systems, such a static warehouse is not available in the same form. Instead, data emerges continuously as a stream during vehicle operation, and only a limited fraction of this stream can be recorded, transmitted, and processed.
As a result, the target dataset is not selected from an already available data repository, but must be generated through the collection process itself. This shifts the problem from post-hoc dataset selection to online data valuation and acquisition. To collect a high-quality dataset under these conditions, it is necessary to clarify which dataset quality characteristics are relevant and which collection strategies are able to realize them under Fast Data constraints.
Accordingly, the problem addressed in this work is how to generate high-quality machine learning datasets from real-world vehicular data streams when exhaustive recording is infeasible. This requires, first, a definition of dataset quality for data collection, second, an analysis of available methods for dataset generation and data collection, and third, the derivation of requirements for Fast Data collection strategies that are dataset-state-aware and structure-aware.

%% file: src/03_datasets.tex
\section{Dataset Quality for Data Collection}

\subsection{Quality of Datasets}
\begin{figure}
    \centering
    \includegraphics[width=1\linewidth]{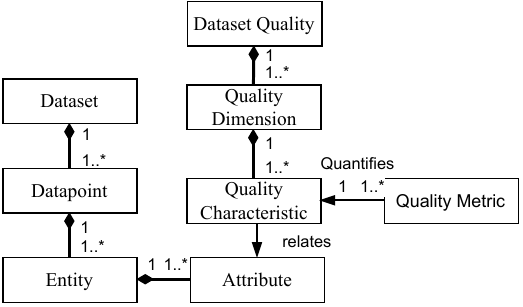}
    \caption{Dataset Quality Model }
    \label{fig:data_set_quality}
\end{figure}
The quality of datasets is a safety-critical aspect for the deployment of \ac{ml} applications in vehicle systems \cite{ISO_8800}. Dataset quality is defined in the ISO 5259 \footnote{Artificial Intelligence – Data quality for analytics and machine learning} as:

    \begin{definition}[Data quality~\cite{ISO_5259-1}]
        \label{def:data_quality}
        Characteristic of data that indicates that the data fulfill the organization's data requirements for a specific context.
    \end{definition}

In \cite{margara_definition_2019,haihua_pdf_2022}, data quality is simplified as \textit{fitness of use}. It refers to the extent to which the available data are suitable for meeting the needs of a specific task.  
The evaluation of data quality is performed at the attribute level using quality characteristics, which are each assigned to higher-level quality dimensions. These quality characteristics are quantified through metrics and describe the data quality within the application context.

\subsection{Quality Dimensions of Datasets} \label{sec:quality_dimension}
The quality of datasets is not a measurable quantity. A systematic understanding is required of how quality in datasets is evaluated (cf. \autoref{fig:data_set_quality}). The literature proposes various dataset evaluation frameworks that analyze datasets from different perspectives and systematize qualitative dimensions (cf. Table~2.2 in~\cite{haihua_pdf_2022}).
A semiotic perspective categorizes the quality of information and data into syntactic, semantic, and pragmatic dimensions~\cite{price_semiotic_2016,ISO_8000}. 
A categorization into intrinsic, contextual, representational, and accessibility dimensions is described in \cite{wang_beyond_1996} and further developed in \cite{zhou_survey_2024,priestley_survey_2023,lee_aimq_2002}. 

In the standards ISO 25012\footnote{Software engineering: Software product Quality Requirements and Evaluation (SQuaRE): Data quality model} and ISO 5259, data quality is divided into \textbf{inherent} and \textbf{system-dependent} aspects \cite{ISO_25012,ISO_5259-1}. The inherent aspect describes the intrinsic potential of data to fulfill a specific task, while the system-dependent dimension characterizes the extent to which this potential can be utilized within a given usage context.  
The categorization of these dimensions is also considered in the context of \ac{ml} and serves as the foundation of this contribution.

\subsection{Quality Characteristics} \label{sec:data_quality_property}
\begin{figure}
    \centering
    \includegraphics[width=1\linewidth]{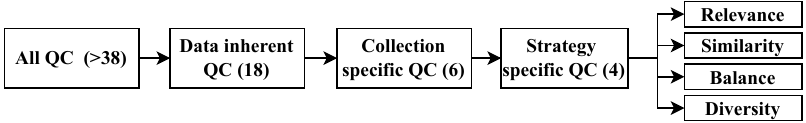}
    \caption{Criteria for the selection of the relevant quality characteristics (QC) for a data collection strategy.}
    \label{fig:qc_selection}
\end{figure}

The dataset evaluation frameworks structure datasets into dimensions, which in turn are characterized by dataset characteristics. A dataset quality characteristic is defined as:
\begin{definition}[Quality characteristic $\mathcal{P}$]
    \label{def:quality_characteristic}
    Category of attributes that affects data quality (\autoref{def:data_quality}) \cite{ISO_5259-1}.
\end{definition}
An attribute is a property or characteristic of an (abstract) object that can be distinguished quantitatively or qualitatively by human or automated means~\cite{ISO_5259-1}.

In contrast to quality dimensions, dataset quality characteristics are measurable. Their definition serves to quantify dataset quality through quality metrics.
Depending on the evaluation framework, the quality characteristics differ in terms of granularity, scope, and number of characteristics considered within the respective categories. 
The relevance of each characteristic depends on the considered dimension and the respective phase in the development life cycle~\cite{gong_survey_2023}.

\subsubsection{Selection of Relevant Characteristics for Data Collection}

The quality evaluation frameworks described in Table~2.2 in~\cite{haihua_pdf_2022}) describe more than 38 quality characteristics whose definitions, scope, and granularity vary depending on the evaluation framework and do not follow a unified taxonomy. In this work, the relevant characteristics are selected using three filtering criteria (cf. \autoref{fig:qc_selection}).

First, only the inherent characteristics are considered, as distinguished from system-dependent characteristics in ISO~5259. Subsequently, a selection is made based on the phases of the data life cycle~\cite{gong_survey_2023,Hutchinson_towards_2021}. For data collection, inherent characteristics are particularly relevant because they are directly influenced by the collection process~\cite{gong_survey_2023,priestley_survey_2023}. 

Finally, a distinction is made according to the data collection strategy. For example, the quality characteristic accuracy is influenced by methods of data compression; however, selection strategies do not affect accuracy.

As a result, the quality characteristics \textbf{relevance}, \textbf{balance}, \textbf{diversity}, and \textbf{similarity} remain. \textbf{Balance} describes the distribution of entities or their attributes within a dataset according to~\cite{ISO_5259-2}. Dataset imbalance can lead to underrepresentation and overrepresentation, which may reduce the performance of \ac{ml} models. Trivial duplication of minority classes does not adequately address this issue and can further impair model performance~\cite{tholke_class_2023,moore_dataset_2023}. \textbf{Diversity} describes the degree to which samples within a dataset differ at the feature level according to~\cite{ISO_5259-2}. Low diversity can increase overfitting and reduce the generalization capability of trained \ac{ml} models~\cite{albattah_impact_2025}. \textbf{Relevance} describes the degree to which a dataset is suitable for a specific target problem according to~\cite{ISO_5259-2}. \textbf{Similarity} describes the degree of resemblance between data points within a dataset according to~\cite{ISO_5259-2}. A high number of highly similar data points can also promote overfitting of the \ac{ml} model. These quality characteristics describe datasets at a conceptual level. Their evaluation requires numerical representation through \textbf{quality metrics}, which map quality characteristics to measurable values.

\subsection{Quality Metrics}
The quality characteristics for data collection introduced in~\autoref{sec:data_quality_property} describe datasets at a conceptual level.  
However, their evaluation requires a numerical representation.  
Quality metrics map quality characteristics to numerical values.
\begin{definition}[Quality metric]
    \label{def:data_quality_metric}
        A quality metric $\mathcal{Q}$ assigns a numerical value to a dataset $\mathcal{D}$ with respect to a characteristic $\mathcal{P}$:
        \begin{equation}
            \mathcal{Q} :\,\,\, (\mathcal{D}, \mathcal{P}) \rightarrow \mathbb{R}.
        \end{equation}
\end{definition}
For each quality characteristic exists multiple metrics like the vendi score~\cite{friedman_vendi_2023} for deversity or k-nearest neigbourg for similarity~\cite{birodkar_semantic_2019}. An evaluation of a different metric and its impact on Data Quality is still future work.

\section{Methods for Dataset Generation} \label{sec:data_gen}
\begin{figure}
    \centering
    \includegraphics[width=0.55\linewidth]{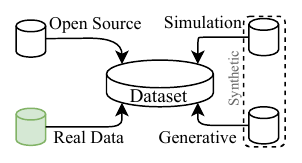}
    \caption{Four sources for generating a dataset.}
    \label{fig:dataset_generation}
\end{figure}
The development of \ac{ml} models is based on datasets that are used for training. These datasets can originate from real-world driving recordings or can be generated synthetically through simulation or generative \ac{ai} models. In addition, publicly available datasets exist, which themselves are based on one of these sources (cf. \autoref{fig:dataset_generation}). 

\subsection{Real-World Driving Data Collection}
Real-world driving data collection records the sensor data stream from the vehicle’s surrounding \ac{odd} driving environment and is commonly practiced in industry for testing and validation~\cite{vw_data_recording}. In addition to real environmental, traffic, and vehicle states, phenomena such as noise and sensor artifacts are recorded, as well as rare events that occur in reality.
With regard to dataset generation, real-world driving provides the ground truth with respect to the quality characteristic of accuracy, as it directly represents the target systems. Consequently, many applications rely on real driving data~\cite{hohl_data_tax_2025}.

\subsection{Synthetic Data Generation}
\subsubsection{Simulation}

Simulation refers to the execution of mathematical models that approximate real-world systems to reproduce their behavior under controlled conditions~\cite{bungartz_modellbildung_2013}. In the context of dataset generation, simulation enables the controllable, scalable, and reproducible creation of synthetic sensor data without physical deployment. It comprises sensor, environment, dynamics, and behavior simulation, covering aspects such as sensor modalities, traffic participant interactions, and environmental conditions (e.g., weather)~\cite{ransiek_adversarial_2025,schulz_fast_2023,schramm_modellbildung_2018}.
Simulation supports dataset generation through three key contributions: dataset augmentation, critical scenario generation, and its use as a validation environment~\cite{padusinski_point_2025,bogdoll_augmenting_2022,klischat_scenario_2020,steinhauser_efficient_2025}.

\subsubsection{Generative AI}
Generative AI for synthetic data generation originated with \textit{Generative Adversarial Networks (GANs)}, which were mainly used for image synthesis but suffer from issues such as \textit{mode collapse}, limiting data diversity~\cite{goodfellow_generative_2014,frid-adar_synthetic_2018,Kossale_mode_collaps_2022}.

More recently, transformer-based \textit{diffusion models} have become dominant, enabling controllable data generation through conditioning (e.g., natural language prompts)~\cite{rombach_high-resolution_2022,feng_survey_2025,hafner_mastering_2024,russell_gaia-2_2025,wang_worlddreamer_2024}. Similar to simulation, they are used for dataset augmentation, critical scenario generation, and validation.

\subsection{Open Datasets}
Open datasets are a key component of data-centric development and provide standardized benchmarks for highly automated driving. They enable the evaluation of methods and \ac{ml} models across perception tasks such as lane detection, object detection, and semantic segmentation~\cite{CurveLane2020,zhang_citypersons_2017,yu_bdd100k_2020,Cordts2016Cityscapes,behley2019iccv}.
Datasets vary in generation method (real, simulated, generative), sensor modality (e.g., camera, LiDAR, multimodal), annotation, and scale. While early datasets were primarily unimodal and camera-based~\cite{geiger_are_2012}, modern datasets are increasingly multimodal, combining multiple sensor types~\cite{nuscenes2019}.
Beyond perception, current datasets address scenario understanding by integrating spatial, semantic, physical, and temporal information, enabling the evaluation of advanced \ac{ml} models in tasks such as visual question answering and decision making~\cite{sohn_framework_2025}. These datasets are often designed for benchmarking rather than training, particularly for assessing \textit{foundation models}, which are trained on large-scale, multi-domain data and evaluated in contexts such as vision-language modeling for automated driving~\cite{yang_qwen3_2025,sohn_framework_2025}.

\subsection{Challenges in Dataset Generation}
\input{fig/generation_comp_table}

The relevance of a dataset always depends on its intended task and research objective, so dataset generation cannot follow a universal notion of quality but must satisfy application-specific requirements~\cite{hohl_data_tax_2025}. For automotive \ac{ml}, these requirements primarily concern diversity, balance, similarity, and relevance, which must be achieved under the constraints of the \textit{Fast Data} paradigm.

ISO~8800 identifies three major factors that limit dataset suitability: \textit{Data Aging Impact}, \textit{Data Reuse Impact}, and \textit{Data-Bias Impact}~\cite{ISO_8800}. Data aging reduces relevance as traffic environments evolve, for example, through new vehicle types, or micro-mobility objects. Data reuse is constrained when existing datasets differ from the target system, such as in sensor setup or calibration. Data bias results from insufficient diversity and imbalance, which is especially critical given the long-tail nature of real-world driving data~\cite{Ackermann_long_tail_2017}.

These limitations affect dataset generation methods in different ways. Simulation and generative methods provide controllability, scalability, and reproducibility, but depend on real-world data for reference or training and remain limited in realism, sensor fidelity, and the representation of rare anomalies~\cite{steinhauser_data_2025,steinhauser_simreal_2024}. Public datasets support efficient benchmarking, but are restricted by aging and reuse effects. Real-world data collection is more expensive and less controllable, yet it offers the highest fidelity, captures genuine anomalous events, and remains essential for both validation and synthetic data generation.

As a result, no single generation method is sufficient for high-quality automotive datasets. Instead, public, synthetic, and real-world data must be combined complementarily. Within this combination, real-world data remains indispensable because it provides the most accurate and current representation of the operational domain and is the ground truth for model training and system validation. Since exhaustive real-world recording is infeasible due to data volume and long-tail effects, \textit{Fast Data} strategies are required to selectively acquire samples that most improve relevance, diversity, and balance.

%% file: fig/generation_comp_table.tex
\begin{table}[t]
\centering
\caption{Comparison of real-world driving, simulation, and generative \ac{ai} across different categories. The evaluation is qualitative and classified as follows: (-): poor, (o): limited, and (+): good.}
\label{tab:vergleich}
\begin{tabular}{lccc}
\toprule
\hspace{10mm}\textbf{Category} & \textbf{Real Data} & \textbf{Sim.} & \textbf{GenAI} \\
\hline
Reproducibility & - & + & + \\
Controlability & - & + &  + \\
{Coverage of rare events} & o & - & - \\
Cost/Scalability & - & +  & + \\
Accuracy & + & o &  - \\
\bottomrule
\end{tabular}
\end{table}

%% file: src/04_strategies.tex
\section{Analysis of Data Collection Strategies}


The generation of high-quality datasets from real-world driving data streams requires the consideration of quality characteristics and thus a targeted selection of data. A data collection strategy is defined as:

\begin{definition}[Data collection strategy]
		\label{def:data_collection_sttrategy}
             Let a data stream $\mathcal{S} = (x_1, x_2, x_3, \dots)$ be given. Then, a data collection strategy is a function $\mathcal{F}$ that generates a dataset $\mathcal{D}$ from the data stream up to a time $T$:
             \begin{equation}
                 \mathcal{F}:\,\,\,\mathcal{S}_\mathrm{T}\to \mathcal{D}.
             \end{equation}
             The quality of the data collection strategy is determined analogously to Definition \ref{def:data_quality_metric} as $\mathcal{Q}:(\mathcal{F}(\mathcal{S}_\mathrm{T}),\mathcal{P})\to\mathbb{R}$:
	\end{definition}

\label{sec:taxonomie_relevante_daten}

\begin{figure}
    \centering
    \includegraphics[width=1\linewidth]{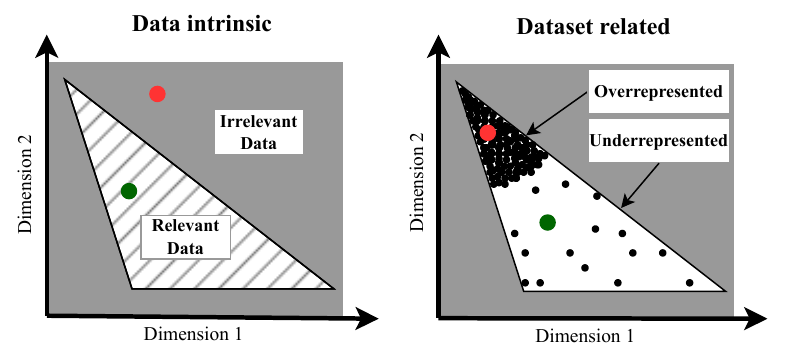}
    \caption{Visualization of the distinction between data-intrinsic (left) and dataset-specific (right) selection criteria in a two-dimensional example.}
    \label{fig:relevance_distribution}
\end{figure}

Data collection strategies can be distingushed into data-intrinsic and dataset-specific selection criteria (cf. \autoref{fig:relevance_distribution}):

\subsection{Data Intrinsic Trigger}

A data-intrinsic trigger function $\mathbf{T}_\mathrm{I}$ is a function that maps a data point from the input space $\Omega_\mathcal{X}$ to a binary outcome, thereby deciding whether an event is triggered:
\begin{equation}
\mathbf{T}_\mathrm{I}:\,\,\,\Omega_\mathcal{X} \rightarrow \{true,false\}.
\end{equation}
Data-intrinsic triggers operate on patterns in data streams and can be categorized into three types:
(1) \textbf{rule-based}, where logical conditions are applied to raw signals (e.g., EDR detecting accidents via thresholds~\cite{edr_2022});
(2) \textbf{semantic}, where human-interpretable concepts from model representations or multimodal embeddings (e.g., detected objects or text-image alignment) are used for trigger decisions~\cite{radford_learning_2021,rigoll_focus_2023,sohn_towards_2025,rigoll_unveiling_2024,xiao_florence-2_2024,rigoll_clipping_2024};
and (3) \textbf{error-based}, where deviations between expected and observed behavior serve as triggers, either explicitly (e.g., OBD following ISO~15031) or implicitly (e.g., shadow mode)~\cite{gyllenhammar_vehicle_2022,gruser_ai-based_2024}.

\subsection{Dataset-Specific Triggers}

A dataset-specific trigger function $\mathbf{T}_\mathrm{D}$ evaluates a data point $x \in \Omega_\mathcal{X}$ in the context of an existing dataset $\mathcal{D}$ and maps it to a real-valued score:
\begin{equation}
\mathbf{T}_\mathrm{D}:\,\,\,\mathcal{D} \times \Omega_\mathcal{X} \rightarrow \mathbb{R}.
\end{equation}
Dataset-specific triggers assess data with respect to dataset properties such as similarity, balance, and diversity. A central category is \textit{anomaly detection}, which targets rare or critical events (e.g., novelties, corner cases, and edge cases) and addresses the long-tail problem in automated driving~\cite{heidecker_application-driven_2021}.

Anomalies can be structured across multiple levels, including pixel, object, domain, scene, and scenario, with dedicated detection methods for each level~\cite{breitenstein_systematization_2020}. Common approaches include statistical methods, clustering, nearest-neighbor techniques, isolation forest, and classification-based models~\cite{hofmockel_anomalieerkennung_2019}, which can further be grouped into prediction-, reconstruction-, generative-, uncertainty-, and feature-based methods~\cite{bogdoll_anomaly_2022}.

However, not all approaches are suitable for online application, highlighting the need for efficient real-time capable methods.

\subsection{Static and Adaptive Definitions of Normality}
\label{sec:normal}

A central aspect of dataset-specific collection strategies is the distinction between \emph{normal} and \emph{abnormal} data. However, the literature defines this distinction inconsistently and at different levels of abstraction. While anomalies are often described explicitly, the corresponding notion of \emph{normality} frequently remains implicit.

In expert-driven approaches, \emph{normal} is specified manually, for example through predefined object classes or environmental conditions~\cite{chan_segmentmeifyoucan_2021,blum_fishyscapes_2021,xue_novel_2019,sakaridis_acdc_2021}. In machine learning-based anomaly detection, \emph{normal} is instead defined implicitly by the training data~\cite{hofmockel_anomalieerkennung_2019}. Although this makes normality dataset-conditioned, it remains static during deployment, since any adaptation requires offline data acquisition, labeling, retraining, and redeployment. Thus, many anomaly-based approaches are only indirectly dataset-specific: they derive their notion of normality from a dataset during development, but operate at runtime with a fixed reference model. In deployment, they therefore resemble data-intrinsic triggers with respect to the live data stream.

In addition, these methods mainly address novelty or out-of-distribution detection and therefore capture deviations from a learned data distribution rather than broader dataset quality objectives such as diversity, balance, or similarity.


Overall, the literature spans static expert-defined or static learned notions of normality. For dataset-specific collection, this distinction is crucial, since static notions only detect deviations from a historical reference. However, adaptive notions enable comparison against the evolving dataset state in real time. This is a prerequisite for closed-loop Fast Data collection.

%% file: src/05_Fast_Data.tex
\section{Towards Closed-Loop Fast Data Collection}
\label{sec:closed_loop_fast_data}

\begin{figure*}
    \centering
    \includegraphics[width=0.85\linewidth]{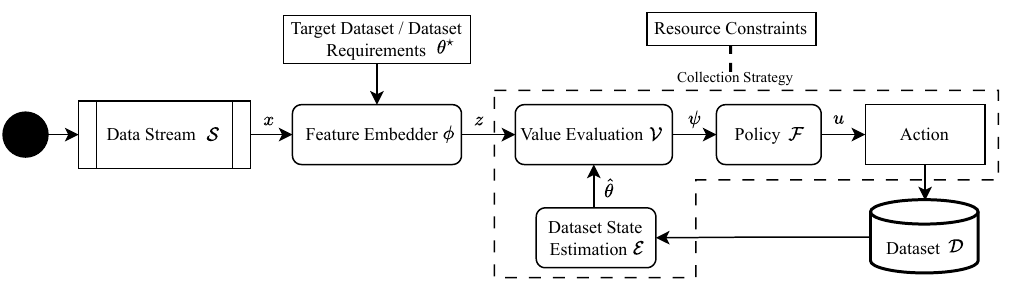}
    \caption{Fast Data collection Paradigm as a closed loop system. }
    \label{fig:closed_loop_fast_data}
\end{figure*}

Existing data collection strategies only partially satisfy the requirements of Fast Data. Data-intrinsic triggers support online decisions at the source, but remain limited to local properties of the current sample. Dataset-specific triggers relate new observations to previously collected data, but in many current systems this relation is established only indirectly through models trained offline on historical data and updated only in delayed retraining cycles. As a result, many state-of-the-art approaches are not dataset-specific during deployment, but merely \emph{dataset-conditioned}: their decision logic depends on a historical dataset, while remaining largely fixed at runtime.

This is insufficient when the objective is not simply to record large volumes of data, but to generate datasets with specific structural properties. Relevance, diversity, balance, and coverage are not pointwise properties of isolated samples. Instead, the utility of a new sample depends on the current state of the collected dataset and on the gap between this state and the desired dataset requirements. Fast Data collection must therefore be formulated as a feedback-driven process in which collection decisions adapt continuously to the evolving dataset.

\subsection{From Open-Loop Recording to Feedback-Driven Collection}
\label{sec:open_to_closed_loop}

Open-loop collection strategies determine what data to record using fixed rules or pre-trained decision functions that do not continuously incorporate feedback from the current dataset. This principle is typical of Big Data pipelines, where data are first recorded and stored, and only later evaluated in centralized infrastructures. Even when anomaly detection is executed online, the underlying decision function is often derived offline and remains unchanged during deployment.

Such strategies are inadequate for controlled dataset generation. First, they cannot account for redundancy, since repeated observations of frequent situations remain indistinguishable from genuinely informative samples. Second, they cannot optimize dataset-level objectives such as diversity, balance, or similarity, because these are relational properties of the dataset as a whole. Third, they cannot adapt collection behavior when the operational domain shifts or when dataset requirements evolve. Open-loop collection may therefore be sufficient for event logging, but not for the structure-aware generation of high-quality machine learning datasets.

\subsection{Formal Model of Closed-Loop Fast Data Collection}
\label{sec:formal_closed_loop}

To make this dependence explicit, we model Fast Data collection as a discrete-time feedback process over dataset state. Let \(x[k]\) denote a datapoint from a data stream~$\mathcal{S}$ at time step \(k\), and let
\[
z[k] = \phi(x[k])
\]
be a feature representation of the current sample. The collected dataset is not represented directly by all stored samples, but by a compact state estimate~\(\hat{\theta}[k]\) that summarizes those dataset properties relevant for collection control. The desired dataset properties are encoded by a target state \(\theta^\star\), which represents the dataset requirements.

The key quantity is the value of an incoming sample relative to the current collection objective. We define this value as
\[
\psi[k] = \mathcal{V}\!\left(\hat{\theta}[k-1], \theta^\star, z[k]\right),
\]
where \(\mathcal{V}\) estimates how strongly the current sample would reduce the mismatch between the present dataset state and the desired target state. In contrast to static anomaly detection, the sample is not evaluated in isolation, but relative to the evolving dataset estimate and the target requirements.

Based on this value, the collection policy determines a control action
\[
u[k] = \mathcal{F}(\psi[k]), \qquad u[k] \in \{0,1\},
\]
 where \(u[k]=1\) means that the sample is collected and \(u[k]=0\) means that it is discarded.

The state estimate is then updated 
\[
\hat{\theta}[k] = \mathcal{E}\!\left(\hat{\theta}[k-1], z[k]\right),
\]
and the dataset evolves as
\[
\mathcal{D}[k] = \mathcal{D}[k-1] \cup \{x[k] \mid u[k]=1\}.
\]

In an open-loop or dataset-conditioned strategy, \(\mathcal{F}\) is fixed after offline training and does not depend on an online-updated estimate \(\hat{\theta}[k]\). In a closed-loop strategy, by contrast, the current dataset estimate directly affects the valuation of future samples and thereby changes subsequent collection decisions. Fast Data collection is therefore \emph{state-aware}, because decisions depend on the current dataset estimate, and \emph{requirement-aware}, because decisions are evaluated against the target state~\(\theta^\star\). An exemplary implementation is given in~\cite{reis2025feedbackcontrolframeworkefficientdataset}.

\subsection{Closed-Loop Interpretation}
\label{sec:control_interpretation}

The formal model admits a control-oriented interpretation. The target state~\(\theta^\star\) acts as the reference variable, the current estimate \(\hat{\theta}[k]\) is the measured dataset state, and the mismatch between both defines the current collection need. The policy~\(\mathcal{F}\) acts as the controller by converting sample value into a collection action. The exogenous stream \(\mathcal{S}\) serves as the source of candidate observations, while the controlled outcome is the evolution of the retained dataset \(\mathcal{D}[k]\).
Importantly, the raw data stream itself is not controlled. What is controlled is the \emph{selection and retention process} operating on that stream. The closed loop therefore acts on dataset formation rather than on data generation. 

\subsection{Architectural Requirements}
\label{sec:closed_loop_requirements_revised}

The formalization above yields a set of requirements for a closed-loop Fast Data collection architecture.

\begin{enumerate}
    \item \textbf{Explicit target state:} The desired dataset properties must be represented explicitly by a target state \(\theta^\star\), encoding objectives such as relevance, diversity, or balance.

    \item \textbf{Online dataset-state estimation:} The system must maintain an updateable estimate \(\hat{\theta}[k]\) of the current dataset state, containing the information needed for collection control.

    \item \textbf{State- and goal-dependent sample valuation:} The utility of an incoming sample must be evaluated relative to both the current dataset state \(\hat{\theta}[k]\) and the target state \(\theta^\star\), not from the sample alone.

    \item \textbf{Realizable control action:} Sample valuation must be translated into a concrete collection action \(u[k]\), such as retain, discard, prioritize, or transmit.

    \item \textbf{Recursive feedback update:} The selected action must update the retained dataset and its state estimate, so that current decisions influence future valuations and the loop is closed.

    \item \textbf{Online source-side execution under resource constraints:} The closed loop must operate near the data source and satisfy Fast Data constraints such as limited time, bandwidth, storage, and compute.
\end{enumerate}

In this perspective, Fast Data collection is not merely online filtering. It is a feedback-driven process for steering dataset formation under explicit objectives and operational constraints. The key shift is therefore not only temporal, namely moving decisions closer to the data source, but structural: dataset generation becomes state-aware, requirement-aware, and continuously adaptive.

%% file: src/06_Outlook.tex
\section{Conclusion and Outlook}
\label{sec:conclusion_outlook}

This paper contributed a structured perspective on Fast Data-driven dataset generation for real-world driving. It introduced a dataset quality model for data collection, compared alternative dataset generation methods, developed a taxonomy of current data collection strategies, and showed that existing approaches are not sufficient for the controlled generation of high-quality datasets because they lack continuous feedback with respect to dataset state and dataset structure. Based on these findings, the paper proposed the Fast Data paradigm as a closed-loop perspective for dataset collection.

The main outcome is a shift from open-loop recording toward state-aware, structure-aware, and requirement-driven data collection under source-side constraints. Future work must now focus on implementing the necessary components of this paradigm and validating their effectiveness in real-world driving applications.